\documentclass[12pt]{article}
\usepackage{epsf,amsmath}
\date{}
\begin{document}

\title{Stringification of Chiral Dynamics: Wess-Zumino interaction
\thanks{Talks given at the 1th
Workshop on
Hadron Structure and QCD, 18-22.May 2004, St.Petersburg, Russia,
at the Bogolubov  Conference on Problems of
Theoretical and Mathematical Physics, 2-6.September 2004, Dubna, Russia
and
at the 6th Conference on
Quark Confinement and the Hadron Spectrum, 21-25.September 2004,
Villasimius, Sardegna, Italy.}}

\author{{\bf A.~A.~Andrianov}$^\dagger$,{\bf
    D.~Espriu}$^\diamond$,
{\bf A.~Prats}$^\diamond$\\
$^\dagger$ V.~A.~Fock Department of Theoretical Physics\\
St. Petersburg State University, Russia\\
Universit\'a di Bologna \& INFN, Bologna, Italy
E-mail: andrianov@bo.infn.it\\
$^\diamond$ Departament d'ECM,
Universitat de Barcelona, Spain\\
E-mail: espriu@ecm.ub.es;\, prats@ecm.ub.es }
\maketitle

\begin{abstract}
\noindent
The QCD hadronic string is supplemented with the reparameterization-invariant boundary interaction
to background chiral fields associated with pions in a way
compatible with the conformal symmetry. It allows the full reconstruction of the P-even part of the Chiral Lagrangian in a good agreement with the phenomenology of P-even meson interactions. The modification of boundary
interaction necessary to induce the parity-odd Chiral Dynamics (WZW action)
is outlined.
\end{abstract}




\section{ Introduction: pion coupling to QCD string}
String description of QCD in the hadronization regime is a
long-standing problem
with a number of theoretical arguments \cite{Nambu,polya} and
phenomenological
evidences \cite{fram,anafesp} as well as with the recent lattice simulations \cite{kuti}
in favor to its viability at intermediate energies (hadron masses).

The crucial low-energy phenomenon in QCD which makes influence on
 Hadron String building for light quarkoniums is
 Chiral Symmetry Breaking. It determines the QCD vacuum and
results in the formation of
light (massless in the chiral limit) pseudoscalar mesons. For the
string dynamics the background chiral fields
$U(x)$ add new couplings \cite{string} involving the string variable $x_\mu(\tau,\sigma)$,
on the
boundary of the string where flavor is attached.
A consistent string propagation in this non-perturbative background has been realized in \cite{string}
where it was adjusted to provide the essential property of string theory - conformal invariance.

The boundary quark fields
$\psi_L(\tau),\psi_R(\tau)$ transform in the fundamental representation
of the light-flavor group $SU(N)$ with $N=2,3$.
The subscripts $L,R$ are related to the
{\it chiral} spinors.
A local hermitian action $S_b = \int d\tau L^{(f)}$ is introduced on the boundary
$ \sigma =0,\, -\infty < \tau < \infty$ to describe
the interaction with background chiral fields
$U(x(\tau)) = \exp(i \pi(x)/f_\pi)$ where $f_\pi\simeq 90 MeV$,
the weak pion decay constant, relates the field $\pi(x)$ to a $\pi$-meson one.

The boundary Lagrangian is chosen to be reparametrization invariant
and in its bare minimal form reads
\begin{eqnarray}
L^{(f)}_{min}=\frac12 i \left(\bar\psi_L U (1 - z) \dot\psi_R  -
\dot{\bar{\psi}}_L U (1 +z)
\psi_R \right) + \, \mbox{\tt h.c. }, \label{lagmin}
\end{eqnarray}where a dot implies a $\tau$ derivative.
It has been proved \cite{string} to provide the E.o.M. of Chiral dynamics and thereby the Chiral
Lagrangian for the
parity-even sector if the conformal symmetry of this boundary
QFT is reproduced, {\it i.e.} the renormgroup
$\beta$ functions of new boundary constants vanish. In particular the dim-4 chiral structural constants \cite{GL} have been
calculated in terms of the product of the Regge trajectory slope $\alpha'
\simeq 0.9$ GeV$^{-2}$,  $f_\pi^2$ and certain rational numbers
(equivalently they can be characterized by the ratio of  $f_\pi^2$ to
the hadron string tension $T = 1/ 2\pi \alpha'$),
\begin{equation}
L_1 = \frac12 L_2 = -\frac14 L_3 = \frac{ f_\pi^2\alpha' }{16}
= \frac{ f_\pi^2}{32\pi T} .
\label{dim4}
\end{equation}
This prediction  is basically supported by ``abnormal'' divergences in
two loops with maximal number of vertices. It fits well the
phenomenological values \cite{expt}.

However the Lagrangian (\ref{lagmin}) is essentially parity even and
thereby
does not contain any vertices which can eventually
entail the anomalous P-odd part of the Chiral Dynamics.
In our talk we outline the
modification of the boundary interaction which might bring the
Wess-Zumino-Witten Chiral action and other parity-odd vertices.

\section{Chiral dynamics on the line}
To approach the required modification we guess on what might be the form of boundary Lagrangian if one
derives it  from the essential part of the Chiral Quark Model projecting it on the string boundary.
The constituent quark fields control properly the chiral symmetry
during the "ein-bein" projection,
$
Q_L \equiv \xi^\dagger \psi_L,\qquad Q_R \equiv \xi\psi_R,\qquad  \xi^2\equiv U$.
 In these variables and in the chiral limit the CQM Lagrangian density and the pertinent E.o.M. read
\begin{equation}
{\cal L}_{CQM} =
i \bar Q \left( \not\!\partial  +  \not\! v +
 g_A \not\! a \gamma_5\right) Q;\quad  i \left( \not\!\partial  +  \not\! v +
 g_A \not\! a \gamma_5\right)Q  = 0,
\label{CQM}
\end{equation}
where
\begin{eqnarray}
 v_\mu \equiv \frac12(\xi^\dagger(\partial_\mu\xi) -
(\partial_\mu\xi) \xi^\dagger),
\qquad
a_\mu \equiv - \frac12 (\xi^\dagger(\partial_\mu\xi) + (\partial_\mu\xi) \xi^\dagger),
\label{va}
\end{eqnarray}and $g_A \equiv 1 - \delta g_A$ is an axial coupling constant of quarks to pions.
We  relegate the effects of constituent quark mass to the gluodynamics encoded in the string interaction.
Then one can decouple the left and right components of boundary fields in  the process
of dim-1 projection.

Let's assume the quark fields to be located on the dim-1 boundary with coordinates
$x_\mu \equiv x_\mu(\tau)$. The first step in
projection of the E.o.M. (\ref{CQM}) can be performed by
their multiplication on $\gamma^\mu \dot x_\mu$ which leads to the following boundary
equations,
\begin{equation}
\left\{ i \left( \partial_\tau  +  \dot x_\mu v^\mu +
 g_A \gamma_5  \dot x_\mu a^\mu  \right)+ \sigma^{\mu\nu}\dot x_\mu
\left(\partial_\nu +  v_\nu +
 g_A \gamma_5 a_\nu  \right)
 \right\}Q =0;\, \sigma^{\mu\nu} \equiv \frac12 i [\gamma^\mu \gamma^\nu] .
\label{projec}
\end{equation}
We notice that this projected Dirac-type equation seems to be associated to the boundary
action with a Lagrangian of type (\ref{lagmin}).

Let us restore the current quark basis of fields $\psi_L$ thereby going back to the original
chiral fields $U$,
\begin{eqnarray}&&\frac12 \left\{ i\left( \{\partial_\tau, U^{\dagger}\}  +
 z  \dot U^{\dagger} \right)+ \sigma^{\mu\nu} \dot x_\mu
\left(\{\partial_\nu, U^{\dagger}\}+
 g_A  \partial_{\nu} U^{\dagger} \right)
 \right\}\psi_L =0;\nonumber\\
&&\frac12 \left\{ i\left( \{\partial_\tau, U\}  +
 z  \dot U \right)+ \sigma^{\mu\nu} \dot x_\mu
\left(\{\partial_\nu, U\}+
 g_A  \partial_{\nu} U \right)
 \right\}\psi_R =0.
\label{projec1}
\end{eqnarray}Now the culminating point of the "ein-bein" projection consists of making the quark fields $\psi$ truly one-dimensional. Namely we define their gradient in terms of the tangent
vector $\dot x_\mu$:
\begin{equation}
\{\partial_\mu, U^{\dagger}\}\psi_L \Rightarrow \frac{\dot x_\mu}{\dot x_\nu\dot x^\nu}
\{\partial_\tau, U^{\dagger}\}\psi_L;\quad
\{\partial_\mu, U\}\psi_R \Rightarrow \frac{\dot x_\mu}{\dot x_\nu\dot x^\nu}
\{\partial_\tau, U\}\psi_R.
\end{equation}
Finally, the projected equations are originated from the boundary Lagrangian,
\begin{eqnarray}L^{(f)} &\equiv& \frac12 i  \left\{\bar\psi_L \left[ \{\partial_\tau, U\}  + \widehat{F}^{\mu\nu} \dot x_\mu
   \partial_{\nu} U\right] \psi_R  + \bar\psi_R \left[ \{\partial_\tau, U^{\dagger}\} - \widehat{F}_\sharp^{\mu\nu} \dot x_\mu
   \partial_{\nu} U^{\dagger}\right]\psi_L \right\};\nonumber\\   \widehat{F}^{\mu\nu} &\equiv& z g^{\mu\nu} + g_\sigma \sigma^{\mu\nu};\quad
  \widehat{F}_\sharp^{\mu\nu} \equiv  \gamma_{0}\left(\widehat{F}^{\mu\nu}\right)^{\dagger}\gamma_{0}, \label{lagfull}
\end{eqnarray}
where, keeping in mind a certain ambiguity in the projection procedure, we  consider  both constants $z$ and $g_\sigma$
as arbitrary ones and search for their values from the consistency of the Hadron string with
chiral fields on its boundary.

\section{ Two-dimensional QCD and beyond}
 The above constructed projection
is unambiguously verified in the two-dimensional version of QCD where the bosonization fixes basic coupling constants in the Chiral Lagrangian.
As in two dimensions  $\gamma_0 = \sigma_1;\, \gamma_1 = - i \sigma_2, \, \gamma_2\,
(\mbox{\it i.e. }\, "\gamma_5") = \sigma_3$  the Lorentz algebra is generated by
$\sigma_{\mu\nu} = i \epsilon_{\mu\nu} \gamma_2$ which must be used in  the boundary action (\ref{lagfull}).

To develop the string perturbation theory we expand the function $U(x)$ in powers of the string coordinate
field $x_\mu(\tau) =x_{0\mu} + \tilde x_\mu(\tau) $, expand the boundary action
in powers of $\tilde x_\mu(\tau)$
and look
for divergences. At one loop one obtains the following condition
to remove the divergences ($\beta$-function$ = 0$ to preserve conformal symmetry ),
\begin{equation}
- \partial_\mu^2 U +\frac12(3+z^2 - g^2_A)\partial_\mu U U^\dagger \partial^\mu U
- i g_A \epsilon_{\mu\nu}\partial^\mu U U^\dagger \partial^\nu U = 0.
\label{anoma}
\end{equation}
Unitarity of chiral fields (= local integrability of Eqs. of Motion (\ref{anoma})) constrains the coupling constants to
$g^2_A - z^2= 1$. The choice in accordance with
the QCD bosonization is $z=0, g_A = 1$. It corresponds to the correct value of the
dim-2 anomaly (last term in (\ref{anoma})). Thus in QCD$_2$ the hadron string induces the WZW action from the vanishing the boundary $\beta$ function already at one-loop level.

In dim-4 QCD the anomaly and the WZW action have dimension 4 and therefore they
are generated by cancellation of  two-loop divergences. The  antisymmetric tensor
$\epsilon_{\mu\nu\rho\lambda}$ in anomalies arises from the well-known algebra of $\sigma_{\mu\nu}$ matrices.
At one-loop level the interplay between coupling constants $z$ and $g_\sigma$
takes place as well with the unitarity condition, $3 g^2_A - z^2= 1$. But now their
 values  are determined from the consistency (local integrability) of the two-loop
equations providing $\beta$-function $ = 0$. As well as for the P-even part of
the Chiral Lagrangian the P-odd anomaly arises from two-loop contributions
with maximal number of vertices and is described in terms of the
product of
 $\alpha'$, $f_\pi^2$ and certain rational numbers. But
now the structural constant of anomalous operator must be quantized in
units of $1/16\pi^2$ \cite{witt}. Therefrom one may arrive to the
relation between  the Regge trajectory slope $\alpha'$ (or the string
tension $T$) and the pion decay constant $f_\pi$, plausibly as
follows, $$f_\pi^2 \simeq \frac{1}{16\pi^2 \alpha'} = \frac{T}{8\pi}.$$
Thus the anomaly unambiguously relates the scales of the Goldstone boson
physics and of the string dynamics.

\section*{Acknowledgments}
One of us (A.A.) is grateful to the Organizing Committees of the 1th
Workshop on
Hadron Structure and QCD, of the Bogolubov  Conference on Problems of
Theoretical and Mathematical Physics and of the 6th Conference on
Quark Confinement and the Hadron Spectrum for financial support. A.A. was also supported by
grant INTAS call-2000 (project 583) and  the Program "Universities of Russia:
Basic Research". The work of D.E was supported by
the EURIDICE Network, grant FPA-2001-3598 and grant 2001SGR-00065.

\end{document}